\documentclass{raa}
\usepackage{graphicx,times}
\usepackage{natbib}
\usepackage{color}
\usepackage{amssymb,amsmath}
\bibpunct{(}{)}{;}{a}{}{,}

\usepackage[a4paper=true,dvipdfm=true,pagebackref=true]{hyperref}
\hypersetup{pdftitle = The title of my PDF, pdfauthor = My name, pdfsubject= The subject, pdfkeywords = keyword1 keyword2 keyword3}
\hypersetup{colorlinks = true, linkcolor = green, anchorcolor = red, citecolor = blue, filecolor = red, pagecolor = red, urlcolor = red}

\begin{document}

   \title{Pulsations of the High-Amplitude $\delta$ Scuti Star YZ Bootis}

 \volnopage{ {\bf XXXX} Vol.\ {\bf X} No. {\bf XX}, 000--000}
   \setcounter{page}{1}

   \author{Tao-Zhi Yang\inst{1,2}, Ali Esamdin\inst{1}, Jian-Ning Fu\inst{1,3}, Hu-Biao Niu\inst{1}, Guo-Jie Feng\inst{1}, Fang-Fang Song\inst{1,2}, Jin-Zhong Liu\inst{1}, Lu Ma\inst{1}
   }

   \institute{ Xinjiang Astronomical Observatory, Chinese Academy of Sciences, Urumqi, Xinjiang 830011, China; \\
    \and
    University of Chinese Academy of Sciences, Beijing 10049, China; {\it yangtaozhi@xao.ac.cn}
    \and
    Department of Astronomy, Beijing Normal University, Beijing 100875, China; \\
\vs \no
 }

\abstract{We present a study on the pulsations of the high-amplitude $\delta$ Scuti star YZ Boo based on photometric observations in Johnson $V$ and $R$ bands with both the Nanshan 1-m telescope of Xinjiang Astronomical Observatory (XAO) and the Xinglong 85-cm telescope of National Astronomical Observatories, Chinese Academy of Sciences (NAOC). The Fourier analysis of the light curves reveals the fundamental radial mode and its five harmonics, with the fourth and the fifth newly detected. 39 new times of light maximum are determined from the light curves, and combined with those in the literature, we construct the $O - C$ diagram and derive a new ephemeris and the determination of a new value of the updated period 0.104091579(2). Besides, the $O - C$ diagram reveals an increasing period change rate of YZ Boo. Theoretical models are calculated and constrained with the observationally determined parameters of YZ Boo. The mass and age of YZ Boo are hence derived as $M$ = 1.61 $\pm$ 0.05 $M_\odot$ and $age$ = (1.44 $\pm$ 0.14) $\times$ 10$^9$ yr, respectively. With both the frequency of the fundamental radial mode and the period change rate, YZ Boo is located at the post-main-sequence stage.
}

   \authorrunning{T.-Z. Yang et al. }            
   \titlerunning{Pulsations of YZ Boo}  
   \maketitle

%
\section{Introduction}           
\label{sect:intro}

$\delta$ Scuti stars are pulsators with short-period, locating inside the classical Cepheid instability strip crossing the main sequence on or above the Hertsprung-Russell (H-R) diagram. With masses between 1.5 and 2.5 $M_{\odot}$, the pulsation periods of these stars are in the range of 18 minutes to 7.2 hours and amplitudes from milli-mag up to several tenths of a mag. The high-amplitude $\delta$ Scuti stars (HADS), as either Population I or II, are found to have one or two dominant radial modes and peak-to-peak light amplitude variations of larger than 0.3 mag. According to the heavy element abundance, stars can be classified as Population I and Population II. Generally, in $\delta$ Scuti stars, Population I stars, as slow rotators with $v sini$ $\leq$ 30 km/s, are young stars and have relatively higher metallicity . They are usually found closer to the main sequence. However, Population II (SX Phoenicis) stars, are usually the metal-poor, more evolved stars and located in global cluster (\citealt{McNamara2000}).

YZ Bootis (=HIP 75373, $\langle V \rangle$ = 10.57 mag, $\mathnormal{P}$$_{0}$ = 0.1041 days, A6-F1), is a high-amplitude $\delta$ Scuti star with a peak-to-peak amplitude of about 0.42 mag (\citealt{Breger1998,zhou06}). Based on the $uvby$$\beta$ photometry, \citet{Joner83} classified YZ Bootis as a Population I star because of the typical $m_{1}$ index of Population I star. This star has a relatively long observational history, dating back to the earlier studies of the star's light variations (\citealt{Eggen1955,Broglia1957,Spinrad1959,Heiser1964,Gieren1974}) and the classification used to be either RR Lyrae or Population I HADS.

On the period variability of YZ Boo, \citet{Broglia1957} revealed that there existed variations in the light curves from night to night and the largest variations of amplitude can be up to 0.07 mag in blue band. They also found that the times of maximum seems to present periodically about $\pm$ 0.0015 days (\citealt{Broglia1957,Heiser1964}). Subsequent observations (\citealt{Szeidl1981,Jiang1985,Peniche1985,Hamdy1986}) seems to indicate YZ Boo has different continuous period increases around a value of (1$\mathnormal{/ P)dP/dt}$ = $\sim$ 3 $\times$ 10$^{-8}$yr$^{-1}$ . Recently, \citet{zhou06} presented an investigation on the stability in period of YZ Boo and claimed that the period change of YZ Boo is still inclusive. \citet{Ward2008} gave a new value of (1$\mathnormal{/ P)dP/dt}$ = 6.3(6) $\times$ 10$^-$$^9$ yr$^{-1}$, assuming this star is increasing and changing smoothly. However, it is still not an easy task to decide whether its period is constant or varying.

The aim of this work is to report a detailed study of the period changes of YZ Boo, mainly using extensive time-series CCD photometry observations from 2008 to 2015 at both the Nanshan Station and Xinglong Station in China. The paper is organized as follows. In Section 2, we describe our new observations and present the data reduction procedure. In Section 3, the analysis of the pulsations of YZ Boo is performed and hence give the corresponding results. We constructed the classical O-C diagram and presented a new ephemeris in Section 4. The theoretical model and the calculation of eigen-frequency and frequency change rate are given in Section 5. Discussion and conclusions are given in the last Section.

\section{Observations and Data Reduction}
\label{sect:Obs}

In order to investigate the variability in period, YZ Boo was observed from 2008 Feb to 2015 May. The CCD images collected between 2008 and 2013 were mainly from the Xinglong 85-cm telescope, and the data from 2014 to 2015 were obtained with Nanshan 1-m telescope (hereafter NOWT). The Xinglong 85-cm is equipped with a PI MicroMax: 1024BFT CCD camera and the field of view is 16.5$^{'}$$\times$16.5$^{'}$, corresponding to an image scale of 0$^{''}$.96 pixel$^{-1}$ (\citealt{zhou2009}). The NOWT  is equipped with a standard Johnson multi-color filter system (i.e. $UBVRI$ filters)(e.g. \citealt{cousins1976}) and an E2V CCD203-82 (blue) mounted on the primary focus. The CCD camera has 4096 $\times$ 4136 pixels, corresponding to a field of view of 1.3 degree $\times$ 1.3 degree  at a focal ratio of 2.159 with a full image size of 49.15 $\times$49.15 mm$^{2}$ (\citealt{song2016}). Table \ref{tab:Journal} list the journal of observations of YZ Boo.

\begin{table}
\begin{center}
  \begin{tabular}{@{}lllll}
    \hline\hline
    Telescope       & Year & Number of images & Filter & Number of hours \\\hline\hline
    Xinglong 85-cm  & 2008 &   5231              &  $V$   & 25.2  \\
     (1024BFT CCD) & 2009 &   2794              &  $V$   & 13.0  \\
                    & 2011 &   565               &  $V$   & 5.8   \\
                    & 2013 &   642               &  $V,R$ & 7.2   \\\hline
    Nanshan 1-m     & 2014 &   685               &  $V,R$ & 24.6  \\
    (E2V CCD203-82) & 2015 &   1968              &  $V,R$ & 52.9  \\\hline
    Total           &      &   11885             &        & 128.7 \\
    \hline\hline
  \end{tabular}
  \caption{ Journal of Photometric Observations of YZ Boo. Xinglong 85-cm stands for the 85-cm telescope at Xinglong station of China and Nanshan 1-m is the Nanshan 1-m wide-field telescope at XAO.}
  \label{tab:Journal}
\end{center}
\end{table}

\begin{figure}[h]
\begin{center}
  \includegraphics[width=0.80\textwidth]{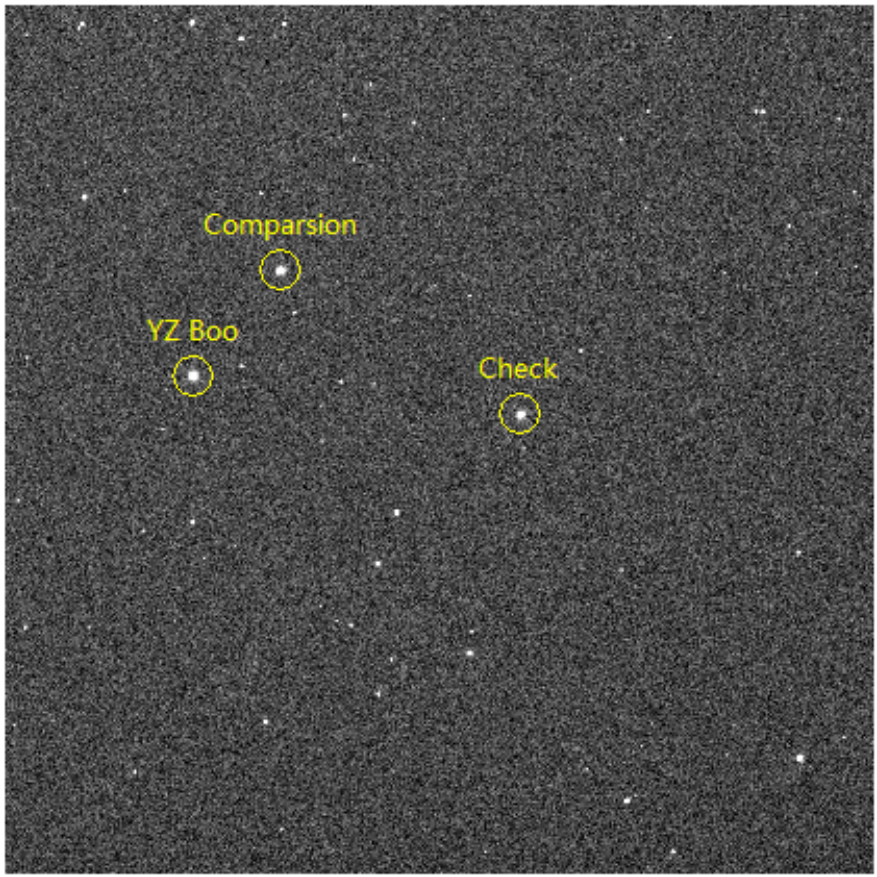}
  \caption{CCD image (18.75$^{'}$$\times$18.75$^{'}$) of YZ Boo
  (RA=15:24:07.0, DEC=36:52:00.6, 2000.0)
  taken with Nanshan 1-m telescope. North is up
  and east is to the left. YZ Boo, the comparison, and the check
  star are marked.}
    \label{fig:YZ_Boo}
\end{center}
\end{figure}

 All the time-series CCD images were reduced with the standard IRAF \footnote{Image Reduction and Analysis Facility,
http://iraf.noao.edu/} routines. Firstly, all the CCD frames were calibrated by removing the bias level using the overscan data, and flat fielding using master flat fields. For the data from NOWT, the dark correction was not considered since the CCD camera was operating at about - 120 $^{\circ}$C with liquid nitrogen cooling, hence the thermionic noise was less than 1 e pix$^{-1}$h$^{-1}$ at the temperature, and at about - 45 $^{\circ}$C for the images from Xinglong 85-cm telescope, so the dark correction was also not considered.  Then, the IRAF APPHOT package was employed to perform aperture photometry. In order to optimize the sizes of the aperture, five to ten different apertures were used for the data in each night and the apertures which presented the minimum variance of magnitude differences of the check star relative to the comparison star were taken. The star C1=GSC 2569-1184 ($V$ = 11.6 mag) was detected as a non-variable within the observational error and then used as the comparison to obtain differential magnitudes for the variable. This comparison star was also used by \cite{Derekas2003}. The star C2 = GSC 2569-1050 ($V$ = 11.4 mag) was used as the check star. Figure \ref{fig:YZ_Boo} shows an image of the field of YZ Boo taken with the NanShan 1-m telescope, on which the variable, the comparison, and the check star are marked. The standard deviations of the differential magnitudes between C2 and C1 yielded an estimate of the mean observational error of about 0.005 mag. In total, we obtained 11885 measurements in $V$ band during 48 nights for YZ Boo. Figure \ref{fig:85cm} and Figure \ref{fig:1m} show the light curves of YZ Boo in Johnson $V$ band observed with Xinglong 85-cm telescope from 2008 to 2013 and NOWT  from 2014 to 2015, respectively.

\section{Pulsation Analysis}

To study the pulsations of YZ Boo, we performed Fourier analysis with all the data in $V$ band using PERIOD04 (\citealt{Lenz2005}), which makes Fourier transformations of the light curves to search for significant peaks in the amplitude spectra. The light curves are fitted with the following formula,

\begin{equation}
m=m_{0} + \Sigma\mathnormal{A}_{i}sin(2\pi(\mathnormal{f}_{i}\mathnormal{t} + \phi_{i}))
\end{equation}

Apart from the fundamental frequency $f$$_0$ and its harmonics 2$f$$_0$, 3$f$$_0$, 4$f$$_0$, as mentioned in \citet{zhou06}, another two harmonics 5$f$$_0$ and 6$f$$_0$ are detected. Generally, it is reasonable that a frequency whose signal-to-noise ratio is lager than 4 (i.e. $S$/$N$ $>$ 4.0) is considered significant (\citealt{Breger1993,Kuschnig1997}). Table \ref{tab:Frequency} lists all the significant frequency solutions including $f_0$ and its 5 harmonics frequencies.

\begin{table}[h]
\begin{center}
  \begin{tabular}{@{}lllll}
    \hline\hline
          & Frequency & Amplitude  & Phase   & S/N \\
          & (c/day)   & (mmag)     & (0 - 1) &      \\\hline
       $\mathnormal{f}$$_0$  &   9.6069   & 191.3  &  0.633   &  177.8 \\
    2  $\mathnormal{f}$$_0$  &   19.2138  & 59.2   &  0.475   &  97.0  \\
    3  $\mathnormal{f}$$_0$  &   28.8207  & 21.9   &  0.765   &  79.2  \\
    4  $\mathnormal{f}$$_0$  &   38.4277  & 9.5    &  0.639   &  33.6  \\
    5  $\mathnormal{f}$$_0$  &   48.0346  & 4.8    &  0.873   &  18.0  \\
    6  $\mathnormal{f}$$_0$  &   57.6435  & 2.2    &  0.869   &   4.6  \\
\hline\hline
  \end{tabular}
  \caption{Results of the frequency analysis of YZ Boo.}
  \label{tab:Frequency}
\end{center}
\end{table}

\begin{figure}
\begin{center}
  \includegraphics[width=0.98\textwidth]{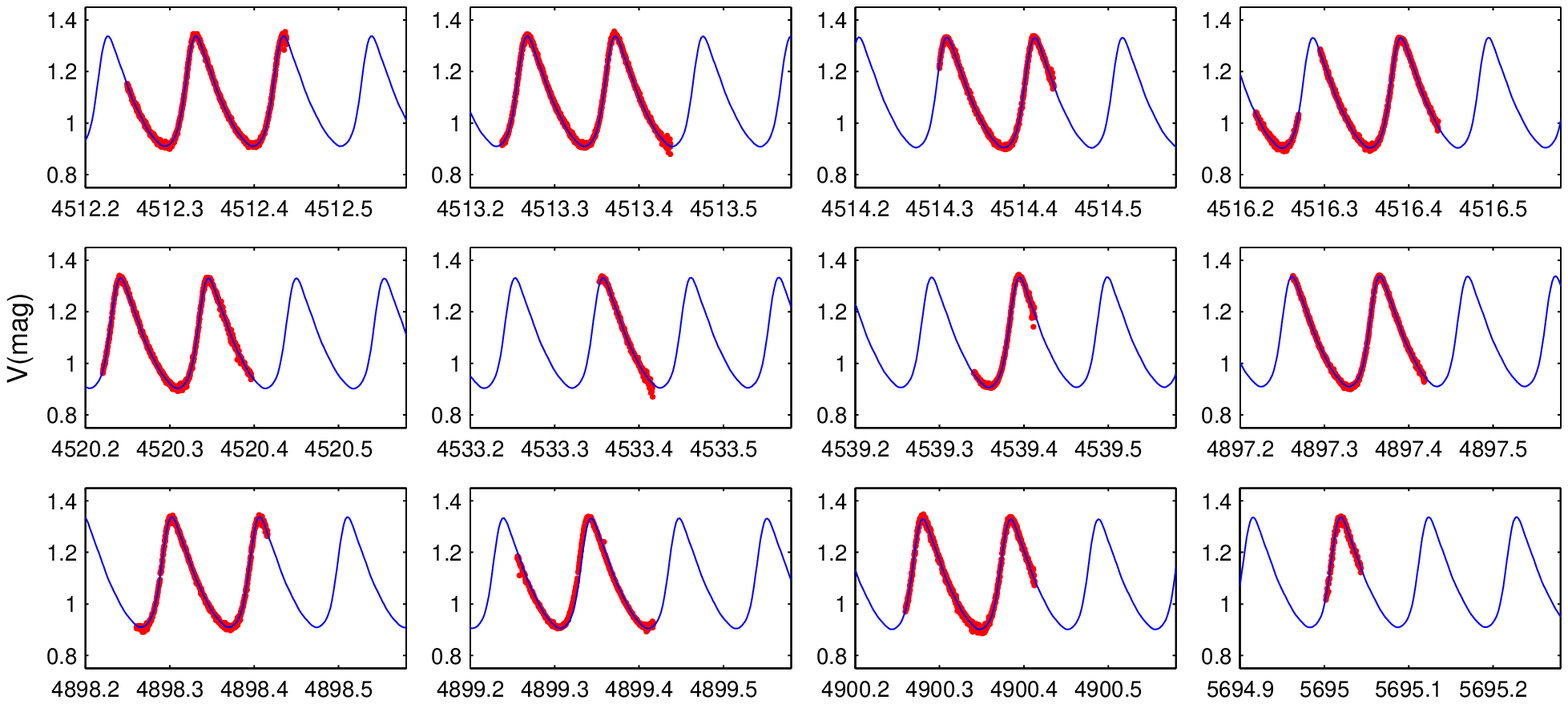}
  \includegraphics[width=0.98\textwidth]{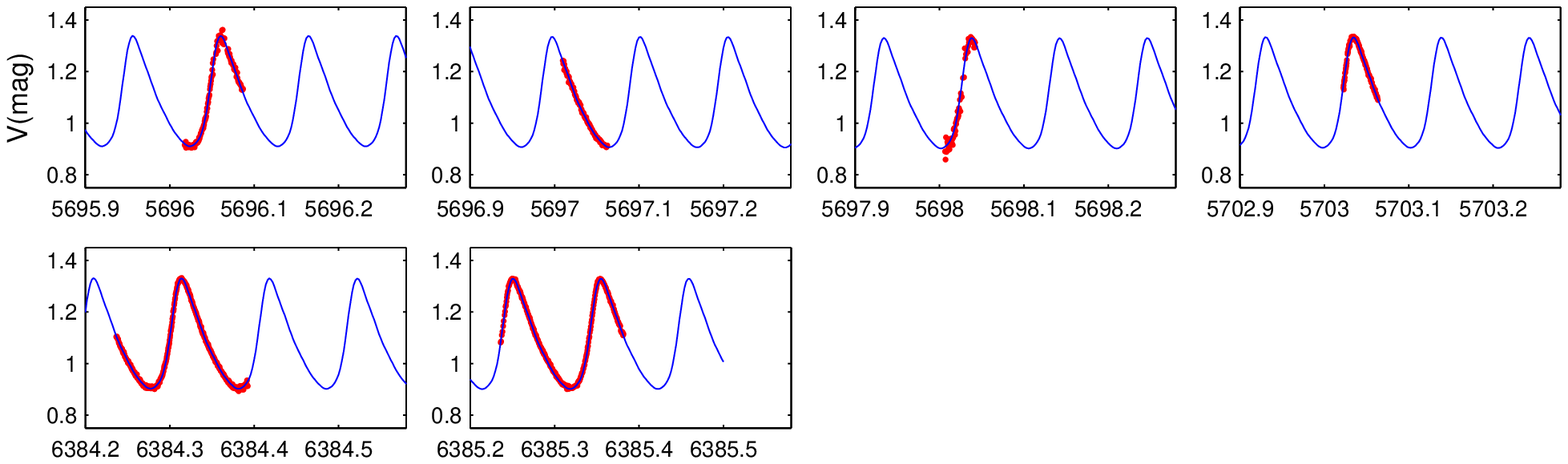}
  \caption{Light curves of YZ Boo in $\mathnormal{V}$ band from 2008
to 2013 observed with Xinglong 85-cm telescope. The solid curves
represent the fitting with the 5 frequency solutions listed in
Table 2.}
  \label{fig:85cm}
\end{center}
\end{figure}

\begin{figure}
\begin{center}
  \includegraphics[width=0.98\textwidth]{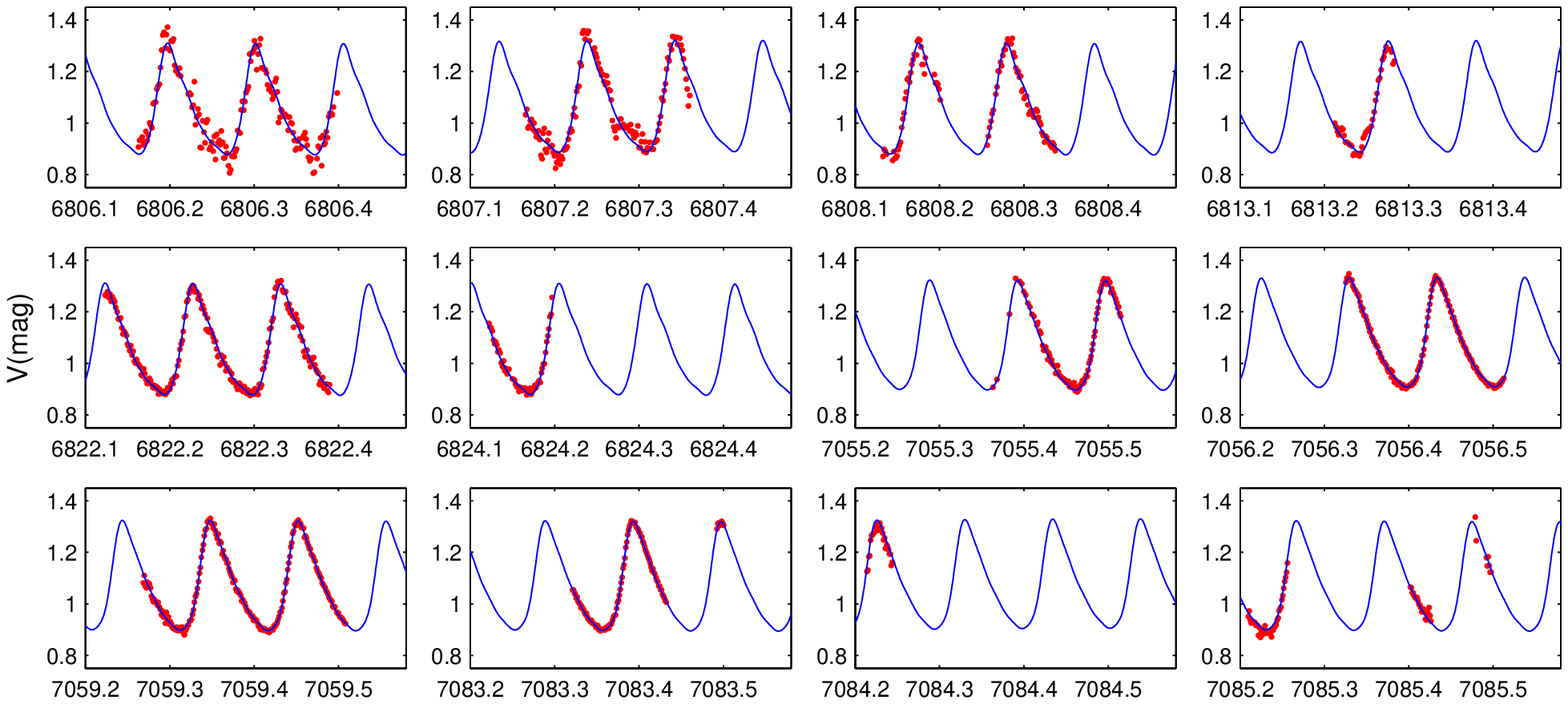}
  \includegraphics[width=0.98\textwidth]{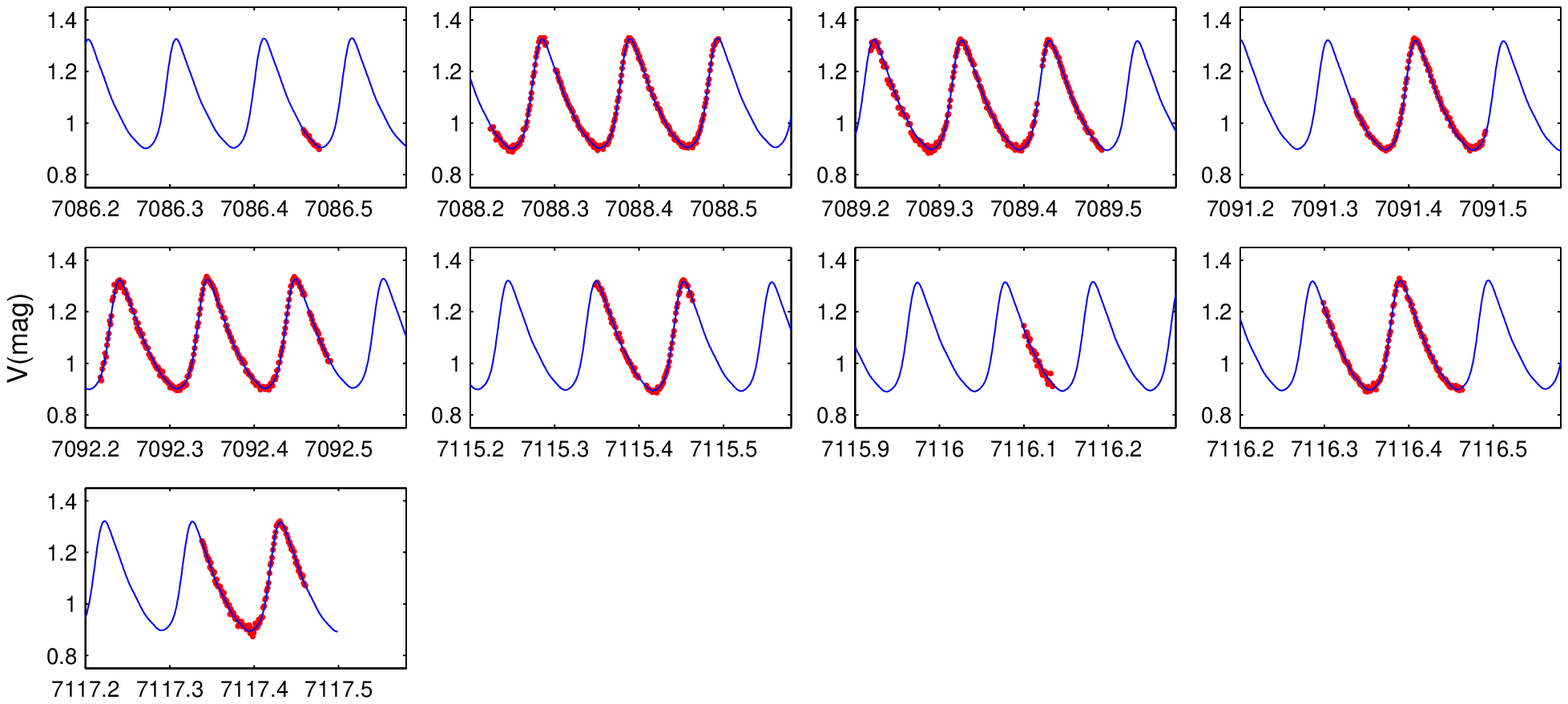}
  \caption{Light curves of YZ Boo in $\mathnormal{V}$ band from 2014
to 2015 observed with the Nanshan 1-m telescope. The solid curves
represent the fitting with the 5 frequency solution listed in Table 2.}
  \label{fig:1m}
\end{center}
\end{figure}

\begin{figure*}
\begin{center}
  \includegraphics[width=0.95\textwidth]{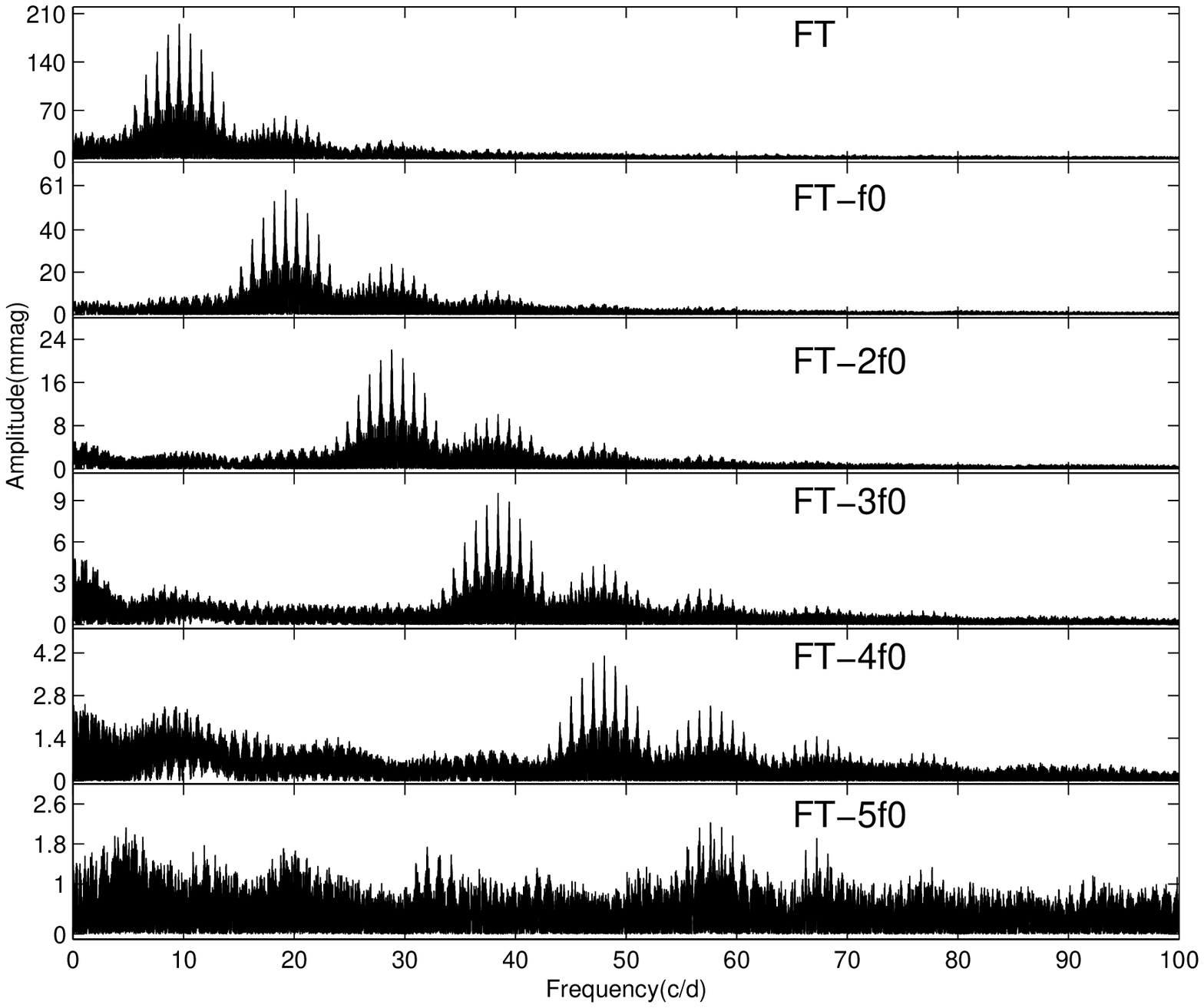}
  \caption{Fourier amplitude spectra of the frequency pre-whitening
   process for the light curves in $\mathnormal{V}$ observed from
   2014 to 2015 with Nanshan 1-m telescope. }
  \label{fig:spectra}
\end{center}
\end{figure*}

 Figure \ref{fig:spectra} shows the amplitude spectra and the pre-whitening procedures for the light curves in $V$ band observed from 2014 to 2015. The residuals of YZ Boo after all the 6 significant frequencies fitting is only 0.0087 mag, which indicate the modelled curves fit the light curves well. One should note that we usually do not consider the peaks in the low-frequency domain (typically in the range of 0-3 cycles day$^{-1}$) as the significant signals of the variable star, because it originate in the instrument sensitivity instability and variations of sky transparency in the low-frequency domain (\citealt{yang2012}).

As can be seen from Figure \ref{fig:85cm} and Figure \ref{fig:1m}, the the modelled curves fit well the observed light curves, which shows that the fundamental frequency and its harmonics can explain the pulsation behavior of YZ Boo. Therefore, YZ Boo is consider as a mono-period variable star since there is only one dependent frequency in the derived frequency.

\section{Times of Maximum Light and $O$ - $C$ diagram }

To examine potential long-term period changes, the classical $O$ - $C$ diagram was constructed.
Firstly, the new times of the light maximum of the light curves were derived. We fitted the light curves around the light maxima using a third or fourth order polynomial. The fitting errors are within the uncertainties that are estimated with Monte Carlo simulations of 200 iterations for each light maximum. The error of maximum determination is from 6 s to 60 s, depending on the data. 39 new maximum times were obtained in $V$ band and listed in Table \ref{tab:Maxtime}.

Secondly, we combined the newly determined times of maximum light with those provided by earlier studies
(\citealt{Eggen1955,Broglia1957,Spinrad1959,Broglia1961,Heiser1964,Fitch1966,Gieren1974,Langfod1976,Rolon1976,Szeidl1981,Joner83,Jiang1985,
Peniche1985,Hamdy1986,Kim1994,Agerer99,Agerer00,Agerer2003,Derekas2003,Jin2003,Hubscher0506,Klingenberg2006,zhou06,Ward2008}).

\begin{table}[h]
\begin{center}
  \caption{Times of light maximum derived from the new light curves.}
  \begin{tabular}{@{}llll}
    \hline\hline
     HJD       & $\sigma$  &    HJD        & $\sigma$\\
   (2450000+)  &           &   (2450000+)  &   \\\hline
 4512.3314     &  0.0001   &  6385.2508    &  0.0002 \\
 4513.2683     &  0.0001   &  6385.3547    &  0.0002 \\
 4513.3716     &  0.0001   &  7055.4971    &  0.0004 \\
 4514.3081     &  0.0001   &  7056.4334    &  0.0002 \\
 4514.4125     &  0.0001   &  7059.3482    &  0.0002 \\
 4516.3906     &  0.0001   &  7059.4520    &  0.0002 \\
 4520.2418     &  0.0001   &  7083.3933    &  0.0001 \\
 4520.3458     &  0.0001   &  7084.2263    &  0.0007 \\
 4539.3942     &  0.0002   &  7088.2864    &  0.0003 \\
 4897.3655     &  0.0001   &  7088.3894    &  0.0002 \\
 4898.3024     &  0.0001   &  7089.3267    &  0.0002 \\
 4898.4061     &  0.0002   &  7089.4303    &  0.0002 \\
 4899.3408     &  0.0001   &  7091.4089    &  0.0002 \\
 4900.2802     &  0.0001   &  7092.2407    &  0.0002 \\
 4900.3843     &  0.0002   &  7092.3456    &  0.0002 \\
 5695.0198     &  0.0002   &  7092.4493    &  0.0002 \\
 5696.0620     &  0.0003   &  7115.4530    &  0.0002 \\
 5698.0375     &  0.0003   &  7116.3902    &  0.0002 \\
 5703.0345     &  0.0001   &  7117.4317    &  0.0002 \\
 6384.3147     &  0.0003   &                         \\\hline\hline
  \end{tabular}
  \label{tab:Maxtime}
\end{center}
\end{table}

A total of 248 times of maximum light were collected. As previous studies, we do not consider the maxima that have been derived from either photographic or visual observations. The seven photometrically determined data points omitted by \citet{zhou06} and \citet{Ward2008} have also been discarded in this study.

To calculate the $O$ - $C$ values and their corresponding cycle numbers, we adopted the revised ephemeris (\citealt{Ward2008}),

\begin{equation}
HJD_{max} = 2442146.3552(2) + 0.104091576(3)\times E
\end{equation}

In this way, the period of YZ Boo is determined as 0.104091579(2), which is near to that of \citet{Ward2008}. The new period result in a more precise linear ephemeris of

\begin{equation}
HJD_{max} = 2442146.3552(2) + 0.104091579(2) \times E
\end{equation}

\begin{figure*}
\begin{center}
  \includegraphics[width=1.00\textwidth]{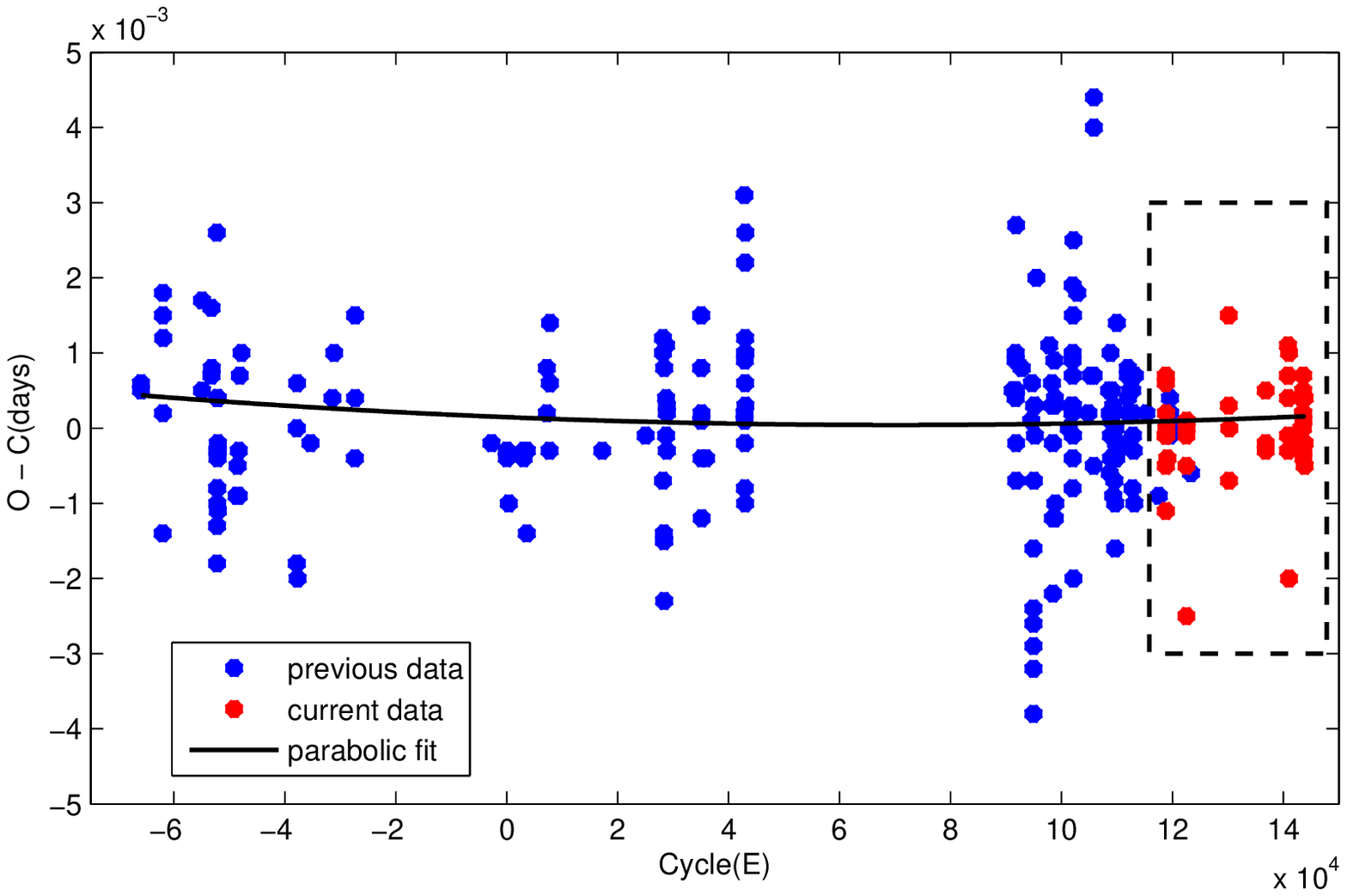}
  \caption{$O$ - $C$ diagram of YZ Boo. The $O$ - $C$ values are in days. Blue dots represent previous 202 points and red dots in the dashed box are new 39 points. The black solid line indicates the parabolic fit concerning a continuously increasing period change.}
    \label{fig:O-C}
\end{center}
\end{figure*}

Figure \ref{fig:O-C} plots the $O$ - $C$ value versus the cycle number of YZ Boo. The standard deviation of the residuals of the linear fit to the $O$ - $C$ values is 0$^d$.0013 .

As the previous studies, we make a parabolic fit to the 241 data points and obtain a continuously changing period. The new ephemeris is,

\begin{equation}
\begin{split}
HJD_{max} = &2442146.3550(2)  + 0.104091570(3) \times E \\+
&1.00(13) \times 10 ^{-13} E ^{2}
\end{split}
\end{equation}

with the standard deviation of the residuals of parabolic fit to the $O$ - $C$ values 0$^d$.0012. The rate of period change is (1$\mathnormal{/ P)dP/dt}$ is derived as 6.7(9) $\times$ 10$^-$$^9$yr$^-$$^1$, which is similar to the results of \citet{zhou06} of 5.0($\pm$3) $\times$ 10$^-$$^9 $yr$^{-1}$, \citet{Ward2008} of 6.3(6) $\times$ 10$^-$$^9$ yr$^{-1}$, \citet{Boonyarak2011} of 5.86 $\times$ 10$^-$$^9$ yr$^{-1}$. As the standard deviations of the residuals of both the liner and the parabolic fits are close to each other, we compared the significance for the fits with these two methods using a two-sided F-test with a 95$\%$ confidence interval. From this test, no significant difference was found in these two fits. However, the parabolic fits was still meaningful, since we can obtained the period change rate of YZ Boo was positive and the order was comparable to the theoretical value. Hence, the two fits are equally acceptable.

\section{Constraints from the theoretical models}

\subsection{Physical Parameters}

We investigate previous studies of YZ Boo to collect its basic physical parameters. Based on $ubvy$$\beta$ photometric observations for YZ Boo, \citet{Joner83} provided some basic information, e.g. the effective temperature $<$$T_{eff}$$>$, average surface gravity $<$log $g$$>$ and its variation range, metal abundance [$Fe/H$] and its bolometric magnitude $M$$_{bol}$. According to the calibration given by \citet{McNamara85}, \citet{McNamara97} provided a new value of metal abundance. They got the values of mass, bolometric magnitude, age, average surface gravity and its variation range. All the parameters of YZ Boo mentioned above are listed in Table \ref{tab:Parameters}.

\begin{table}[h]
\begin{center}
  \caption{Physical parameters of YZ Boo. }
  \begin{tabular}{@{}llll}
    \hline\hline
   Parameters  &  Joner et al. &    McNamara &   Interval \\
               &  (1983)             &    (1997)   &            \\\hline
[$Fe/H$]      & -0.25 $\pm$ 0.20 & -0.60 $\pm$ 0.20 & [-0.80, -0.05] \\
$M_{bol}$     & +1.7  &  1.59  & [+1.4, +2.0] \\
log $g$       & 3.81 - 4.05 & 3.83 - 3.90 & [3.83, 4.02] \\
log($L$/$L$$_{\odot}$) & - & - & [1.1, 1.34] \\
$T_{eff}(K)$   & 7590 & 7490 & [7310, 8130]  \\
$M/M_{\odot}$  &  -   & 1.57 &  -            \\
Age(yrs)       &  -   & 1.58 $\times$ 10$^9$  & - \\
  \hline\hline
  \end{tabular}
  \label{tab:Parameters}
\end{center}
\end{table}

\subsection{Constraints from $f_0$}

Modules for Experiments in Stellar Astrophysics (MESA) is a group of source-open, powerful, efficient and thread-safe libraries for a wide range of applications in computational stellar astrophysics \citep{Paxton2011}. A one-dimension stellar evolution module, MESA star, combines many of the numerical and physical modules for simulations of a large number of stellar evolution scenarios ranging from very-low mass to massive stars, including advanced evolutionary phases. Based on the adiabatic code ADIPLS (\citealt{Christensen08}), the "astero" extension to MESAstar provides an integrated method that passes results automatically between MESAstar and the new MESA module \citep{Paxton2013}.

To obtain more precise values of the physical parameters, we calculated models with different masses and metal abundances. We use the formula [$Fe/H$] = $log(Z/X)$ - $log(Z/X)$$_\odot$ and the formula $X + Y + Z = 1$ to calculate the initial $Z$.  Table \ref{tab:Initial-grid} list the intervals of parameters and the step in calculations of models. In all the calculations, we fixed the mixing-length parameter $\alpha$$_{MLT}$ = 1.89, as the choice has a very small effect on our theoretical models (\citealt{yang2012}). Besides, the convective overshooting parameter $f_{ov}$ = 0.015 was taken as the initial value of MESA.

\begin{table}
\begin{center}
  \caption{The gridpoints of the model.}
  \begin{tabular}{@{}llll}
    \hline\hline
   Parameters &    Interval       &  Step    \\\hline
initial $Z$  &  [0.0025, 0.0150] &  0.0025  \\
$M$          &  [1.40, 1.80]     &  0.02    \\
    \hline\hline
  \end{tabular}
  \label{tab:Initial-grid}
\end{center}
\end{table}

As the result, we found the models with the pulsation frequency of the fundamental radial mode, $f_{0}$ = 9.6069 d$^{-1}$ along with the stellar evolution tracks within the constraints of the photometric data from \citet{Joner83}, and obtained the interval of the parameters as listed in Table \ref{tab:Compact-parameters}.

\begin{table}
\begin{center}
  \caption{The parameters determined from the constraint of $f_{0}$}
  \begin{tabular}{@{}llll}
    \hline\hline
   Parameters & Minimum & Maximum  \\\hline
initial $Z$  & 0.0075 &  0.0150    \\
        $M$  &  1.50  &  1.70     \\
    \hline\hline
  \end{tabular}
  \label{tab:Compact-parameters}
\end{center}
\end{table}

\subsection{Constraints from the Period Variations}

The period variation rate of YZ Boo shows a positive change based on a long interval of observations. From a theoretical point of view, the period changes caused by the stellar evolution in and cross the lower instability strip permit an observational test of the stellar evolution theory, provided that other physical reasons for period changes can be neglected (\citealt{Breger1998}).

As indicated by \citet{Breger1998}, the HADS lie at the intersection of the main sequence and the classical instability strip on the H-R diagram. Consequently, we construct the evolutionary models from zero-age-main-sequence and then make it evolve to the end of the post main sequence. As mentioned above, the same values of $\alpha$$_{MLT}$ and $f$$_{ov}$ were adopted, and the effect of rotation was not considered since YZ Boo is a low-speed rotator with a total velocity of 35 $km/s$ (cf. \citealt{Joner83}).

\begin{figure*}
\begin{center}
  \includegraphics[width=0.7\textwidth,height=0.5\textwidth]{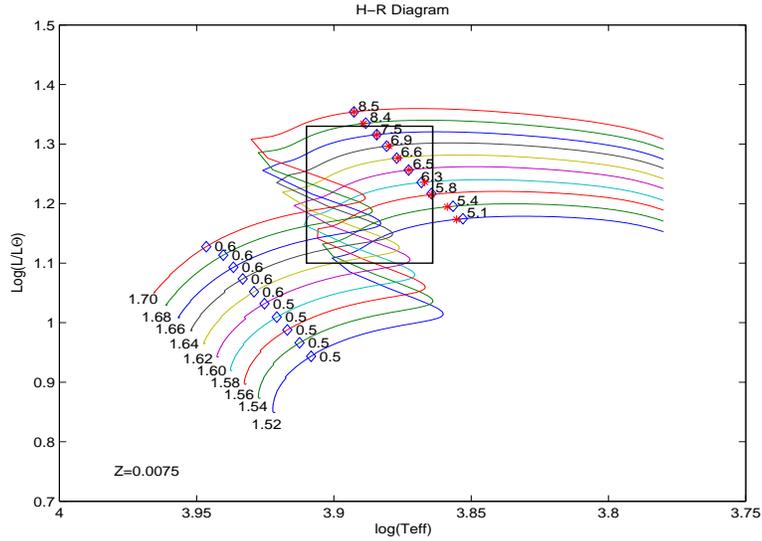}
  \caption{Evolutionary tracks of models with $Z$ = 0.0075 for the mass from 1.52 $M_\odot$ to 1.70 $M_\odot$. The rectangle is determined from the observed parameters of YZ Boo. Diamonds on the tracks indicate the rates of period variations in unit of 10$^-$$^9$yr$^-$$^1$. The asterisks indicate he location of models with $f_{0}$ = 9.6069 d$^{-1}$.}
    \label{fig:HR_3}
\end{center}
\end{figure*}

The evolutionary tracks constructed with $Z$ = 0.0075 for the mass from 1.52 $M$$_\odot$ to 1.70 $M$$_\odot$ is shown in Figures \ref{fig:HR_3}.

The rates of period variations of individual models are also estimated by calculating the slopes of frequencies of the adjecent models along the evolutionary tracks versus the corresponding ages. The frequency differences divided by the corresponding time intervals are taken as the rates of period variations and marked along the evolutionary tracks on Figures \ref{fig:HR_3}.

From the comparison of both the calculated frequencies of the fundamental radial modes of the models with the observed frequencies of YZ Boo, and the theoretical period change rates and the observation determined value of YZ Boo, one can obtain: (1) the evolutionary mass of YZ Boo is $M$ = 1.61 $\pm$ 0.05 $M$$_\odot$, (2) The age of YZ Boo is between 1.30 $\times$ 10$^9$ yr to 1.58 $\times$ 10$^9$ yr, (3)While the metal abundance [$Fe/H$] is about -0.43.

\section{Discussion and conclusions}

With the photometric data observed between 2008 and 2015 from both the Xinglong Station of NAOC and Nanshan Station of XAO, we analysed the pulsations of YZ Boo, and extracted five frequencies, including its fundamental frequency of $f$$_{0}$ = 9.6069 d$^{-1}$ and its 5 harmonics, 2 of which is newly detected. There is no additional frequency found in the residual spectrum after removing these five frequencies. The theoretical light variations of YZ Boo are produced.

The $O$ - $C$ diagram was constructed with 248 times of maximum light either determined from our new observations or collected from the literature, leading to determination of the updated pulsation period of 0.104091579(2) days. Besides, a new ephemeris with a quadratic solution hints a continuously increasing period change of YZ Boo of (1$\mathnormal{/ P)dP/dt}$ = 6.7(9) $\times$ 10$^{-9}$ yr$^{-1}$.
It is consistent with the value predicted from our newly calculated stellar models with masses between 1.4 and 1.8 $M$$_\odot$. The mass of YZ Boo is then determined as $M$ = 1.61 $\pm$ 0.05 $M$$_\odot$ and the location of this variable star on the H-R diagram is limited on a post-main-sequence of the evolutionary tracks.

More observations, especially from multi-site campaigns, would help us to search for more potential pulsation frequencies of YZ Boo and to provide clues to interpret the observed period change rates.


\normalem
\begin{acknowledgements}
Tao-Zhi Yang and Ali Esamdin acknowledge the support from the National Natural Science Foundation of China (NSFC) through grant No. 11273051. This work is partially supported by the CAS "Light of West China" program (2015-XBQN-A-02)" and supported by the Strategic Priority Research Program of the Chinese Academy of Sciences£¬Grant No. XDB23040100. JNF acknowledges the support from the Joint Fund of Astronomy of National Natural Science Foundation of China (NSFC)
and Chinese Academy of Sciences through the grant U1231202, the NSFC grant 11673003, the National Basic Research Program of China (973 Program 2014CB845700 and 2013CB834900), and the LAMOST FELLOWSHIP supported by Special Funding for Advanced Users, budgeted and administrated by Center for Astronomical Mega-Science, Chinese Academy of Sciences (CAMS). GJF acknowledge the support from the National Natural Science Foundation of China (NSFC) through grant NO. 11403088.

\end{acknowledgements}

\bibliographystyle{raa}
\bibliography{bibtex}

\end{document}